%% file: main.tex
\documentclass[conference]{IEEEtran}
\usepackage{booktabs} % For formal tables
\usepackage{color}
\usepackage[bookmarks=false]{hyperref}

\usepackage{amsmath,amssymb,amsfonts}
\usepackage{algorithmic}
\usepackage{graphicx}
\usepackage{textcomp}
\include{my_macro}

\begin{document}

\title{A Structured Learning Approach with Neural Conditional Random Fields for Sleep Staging}%: A Neural Conditional Random Field Approach}

\author
{\IEEEauthorblockN{Karan Aggarwal, Swaraj Khadanga}
\IEEEauthorblockA{University of Minnesota\\
%University\\
Minneapolis, MN\\
\{aggar081, khada004\}@umn.edu }
\and
\IEEEauthorblockN{Shafiq Joty}
\IEEEauthorblockA{Nanyang Technological University\\
Singapore\\
srjoty@ntu.edu.sg}
\and
\IEEEauthorblockN{Louis Kazaglis}
\IEEEauthorblockA{Fairview Health\\
Minneapolis, MN\\
lkazagl1@fairview.org}
\and
\IEEEauthorblockN{Jaideep Srivastava}
\IEEEauthorblockA{University of Minnesota\\
Minneapolis, MN\\
srivasta@umn.edu}

}
\maketitle

\begin{abstract}

Sleep plays a vital role in human health, both mental and physical. Sleep disorders like sleep apnea are increasing in prevalence, with the rapid increase in factors like obesity.  Sleep apnea is most commonly treated with Continuous Positive Air Pressure (CPAP) therapy. 
%which maintains the appropriate pressure to ensure continuous airflow. 
%It is widely accepted that in addition to preventing air passage collapse, increase in deep and REM sleep stages would be good metrics for how well the CPAP therapy is working in improving sleep health. 
Presently, however, there is no mechanism to monitor a patient's progress with CPAP. Accurate detection of sleep stages from CPAP flow signal is crucial for such a mechanism.  We propose, for the first time, an automated sleep staging model based only on the flow signal.
%Developing an automated mechanism that addresses the drawbacks of the manual labeling is the primary goal of this research.

Deep neural networks have recently shown high accuracy on sleep staging by eliminating handcrafted features. However, these methods focus exclusively on extracting informative features from the input signal, without paying much attention to the dynamics of sleep stages in the output sequence.  We propose an end-to-end framework that uses a combination of deep convolution and recurrent neural networks to extract high-level features from raw flow signal with a structured output layer based on a conditional random field to model the temporal transition structure of the sleep stages. We improve upon the previous methods by 10\% 
using our model, that can be augmented to the previous sleep staging deep learning methods. We also show that our method can be used to accurately track sleep metrics like sleep efficiency calculated from sleep stages that can be deployed for monitoring the response of CPAP therapy on sleep apnea patients. Apart from the technical contributions, we expect this study to motivate new research questions in sleep science.% especially towards the understanding of sleep architecture trajectory among patients under CPAP therapy.
%Prior studies on sleep staging have built machine learning models from signals like ECG, actigraphy, or RF sensors.   

%Prior deep learning based works on sleep staging have entirely focused on the input feature extraction, ignoring the dynamics in the output state - transition structure of the sleep stages. 

%We present an end-to-end conditional structured model based on the conditional random field with the deep neural network.  

\end{abstract}


\section{Introduction}
\label{sec:intro}
\input{introduction}

\section{Related Work}
\label{sec:related}
\input{related}

%\section{Problem Statement}
%\label{sec:ps}
%\input{problem}

\section{Model}
\label{sec:model}
\input{model}

\section{Experiments}
\label{sec:experiments}
\input{experiments}

\section{Results and Analysis}
\label{sec:results}
\input{results}

\section{Conclusions}
\label{sec:conc}
\input{conclusions}

\balance
\bibliographystyle{IEEEtran}
\bibliography{IEEEabrv,main}

\end{document}

%% file: my_macro.tex
\usepackage{tikz}
\usetikzlibrary{arrows}
\usepackage{xspace}
\usepackage{latexsym}
\usepackage{amsfonts}
\usepackage{amsmath}
\usepackage{amssymb}
\usepackage{color}
\usepackage{colortbl}
\usepackage{epsfig}
\usepackage{graphicx}
\usepackage{pbox}
\usepackage{paralist}
\usepackage{enumerate}
\usepackage[color,matrix,arrow,all]{xy}
\usepackage{comment}
\usepackage{booktabs}
\usepackage{balance}
\usepackage{stmaryrd}
\usepackage{pifont}
\usepackage{hhline}
\usepackage{listings}
\usepackage{array}
\usepackage{float}
\usepackage[flushleft]{threeparttable}
\usepackage{graphicx,wrapfig,lipsum}
\usepackage{subcaption}
\usetikzlibrary{arrows}
\usepackage{graphicx,dblfloatfix}

\DeclareMathAlphabet{\pazocal}{OMS}{zplm}{m}{n}

\usepackage{array}
\newcolumntype{M}{>{\begin{varwidth}{2cm}}l<{\end{varwidth}}}
\newcolumntype{L}[1]{>{\raggedright\let\newline\\\arraybackslash\hspace{0pt}}m{#1}}
\newcolumntype{C}[1]{>{\centering\let\newline\\\arraybackslash\hspace{0pt}}m{#1}}
\newcolumntype{R}[1]{>{\raggedleft\let\newline\\\arraybackslash\hspace{0pt}}m{#1}}

\newcommand{\ie}{{\em i.e.,}\xspace}
\newcommand{\eg}{{\em e.g.,}\xspace}

 \usepackage[framemethod=TikZ]{mdframed}
\usetikzlibrary{shadows}
\mdfdefinestyle{myframe}{%
    linecolor=black,
    outerlinewidth=0.6pt,
    roundcorner=10pt,
    innertopmargin=10pt,
    innerbottommargin=10pt,
    innerrightmargin=10pt,
    innerleftmargin=10pt,
    skipabove=0.6\baselineskip,
    }

\newcommand{\Ni}{({\em i})~}
\newcommand{\Nii}{({\em ii})~}
\newcommand{\Niii}{({\em iii})~}

\newcommand{\Na}{({\em a})~}
\newcommand{\Nb}{({\em b})~}

\newcommand{\Ls}{\mathcal{L}}

\newcommand{\Is}{\pazocal{I}}

\newcommand\norm[1]{\left\lVert#1\right\rVert}

%% file: introduction.tex
Sleep plays a fundamental role in the physical and emotional recovery of the human body. Sleep deprivation or poor quality of sleep adversely affect the quality of life. %~\cite{mcclain2014associations}. 
Outside of the wake state, sleep can be divided into four stages: \emph{Rapid Eye Movement} (REM), and \emph{Non-REM} (NREM) stages 1, 2, and 3~\cite{silber2007visual}.  
Due to transitory nature of NREM stage 1, stages 1 and 2 are often grouped and classified as light sleep, as compared to deep sleep for NREM stage 3.  Each stage has its role in the recovery process, \eg REM sleep helps in memory consolidation and emotion regulation while deep sleep helps with physical recovery processes. Understanding of a subject's sleep states and their dynamics is necessary for identifying and monitoring various sleep-related 
disorders such as sleep apnea. 

The economic cost of sleep-related disorders is enormous~\cite{shelgikar2014multidisciplinary}.  One of the leading cost burdens is due to \emph{Obstructive Sleep Apnea} (OSA).  OSA is a disorder in which airway collapses during inhalation resulting %in an apnea event, derived from the Greek word for ``without breath". Apneas lead
in a reduced oxygen supply to the brain forcing the patient to wake, causing interrupted sleep.  OSA poses a severe risk, %or in extreme cases, life-threatening consequences, 
\eg OSA is associated with higher rates of heart attacks.  Despite the severity of the condition, it is a mostly undiagnosed disease with an estimated 5\%-20\% prevalence rates among the population~\cite{peppard2013increased} with an estimated cost burden of \$150 billion per year in the US alone~\cite{watson2016health}. 

\begin{figure}[t!]
\centering
\includegraphics[width=0.5\textwidth]{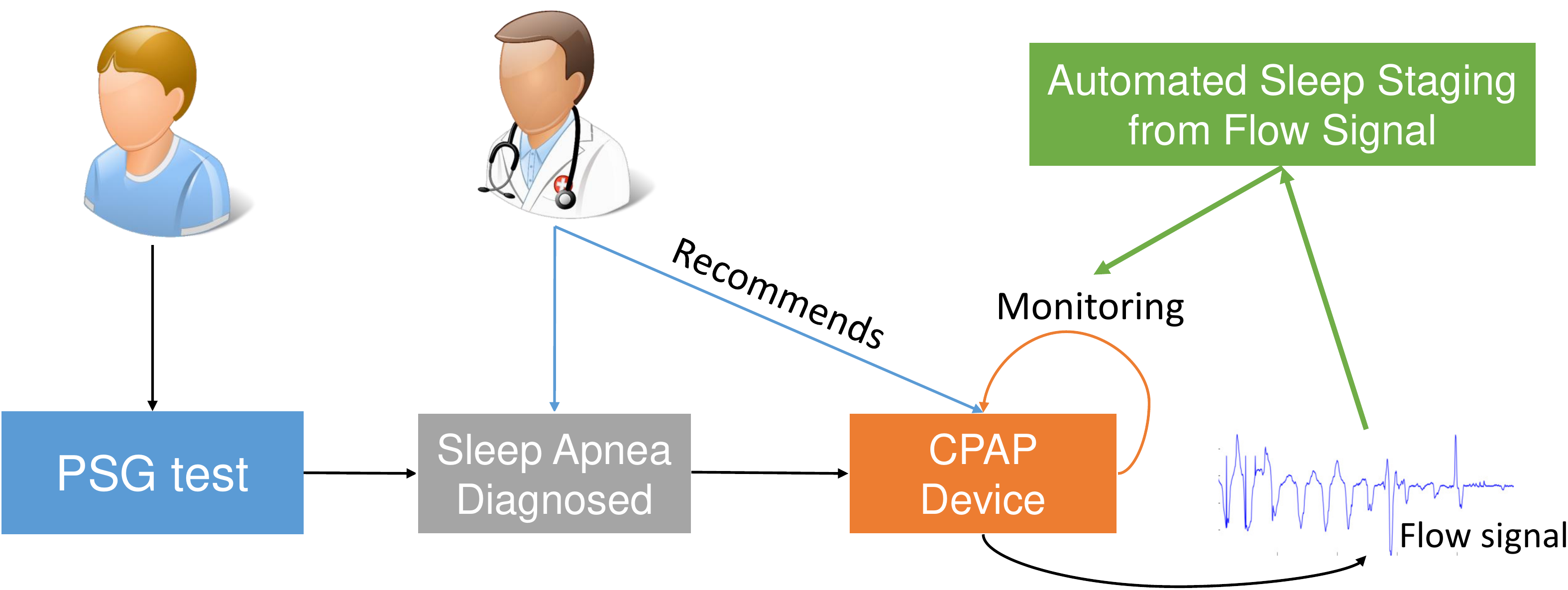}
\caption{An application use case of our model. A patient undergoes Polysomnography (PSG) to ascertain the sleep disorders and is diagnosed with Sleep Apnea. Healthcare provider recommends the CPAP therapy that involves a CPAP device. Flow signal can be obtained from the device daily for monitoring purposes. By adding the automated sleep staging step, we can help the healthcare provider with a means for continuous monitoring of the patient.}
\label{fig:cpapflow}
\vspace{-2em}
\end{figure}

The most prevalent and effective treatment for OSA is \emph{Continuous Positive Airway Pressure} (CPAP) therapy. 
In a CPAP therapy, a user wears a mask, connected to a flow generating device, which delivers an adaptive pressure to prevent the collapse of the airway and track signals like
daily airflow pressure (\emph{flow signal}) data.%, unexpected system leaks, and resistance to airflow. 
This data contains valuable information transmitted to health-care professionals for monitoring the subject's respiratory patterns. However, it is not being utilized actively to monitor the efficacy of patient therapy or sleep quality which in-turn could pave way for an intervention as done in other health-care areas~\cite{yadav2018mining}.  %Sleep apnea leads to disruption of normal sleep architecture, which the treatment of OSA seeks to restore.  %Return of normalcy to sleep architecture is linked to CPAP therapy effectiveness, improved daytime function, and decreased the risk of OSA-related diseases~\cite{verma2001slow}.
The key to  measure the effectiveness of CPAP therapy is to assess the sleep quality by determining the sleep stages. However, to the best of our knowledge, this is the first attempt at determining sleep stages from CPAP-available signals. %have not been reported, nor have there been attempts to determine sleep stages automatically from the same signals. 
Determination of sleep stages has been typically performed on data obtained from Polysomnograms (PSG), which involves an overnight measurement of a variety of biological signals during sleep. The gold standard for securing sleep stages is for trained sleep experts to manually annotate PSG data, a tedium-filled expensive task at best.

Prior studies on sleep staging have focused on 
automating the annotations by using reduced number of sensors from PSG including Electroencephalography (EEG)~\cite{lajnef2015learning,tsinalis2016automatic} or using other more comfortable 
devices like actigraphy~\cite{mantua2016reliability}, cardio-respiratory sensors~\cite{redmond2007sleep}, or no-contact 
sensors~\cite{zhao2017learning}. However, all of these approaches do not have a direct use case - they require additional devices to provide data for sleep staging.  In this work, \emph{we make 
the first attempt to use the CPAP-available flow signal to identify sleep stages automatically}.  CPAP users can know about their sleep health by learning about their sleep states, while the health-care providers can track longitudinal sleep health and overall success of CPAP therapy. Figure~\ref{fig:cpapflow} shows a schematic of the application of our work. A benefit of this study would be to interest the sleep research community in investigating the effect of CPAP therapy on sleep architecture trajectory of OSA patients.

On the technical front, most previous approaches have used hand-crafted features~\cite{lajnef2015learning,tataraidze2016sleep} for the task. Recently, deep neural networks~\cite{tsinalis2016deep,chambon2017deep,mikkelsen2018personalizing} 
have been used for end-to-end learning without manual feature engineering mainly based on convolutional neural networks (CNN). Hybrid recurrent-
convolutional neural networks (R-CNN)~\cite{biswal2017sleepnet,supratak2017deepsleepnet} methods that use CNN as base network fed to the recurrent networks have shown human-expert 
level accuracy on PSG. Adversarial training with R-CNN proposed by Zhao et al.~\cite{zhao2017learning} has shown state-of-the-art results on RF-signals. These methods focus \emph{exclusively  on learning informative abstract features} from the input signal making predictions at each time step independent of the previous sleep state. However, sleep states have a strong transition structure~\cite{kim2009markov}. By not taking into account the dynamics of the sleep states, the deep learning methods have missed out on an essential source of information.

In this work, we propose a new neural network architecture based on \emph{chain-structured conditional random field (CRF)} that explicitly models the temporal dynamics in the sleep states, over the \emph{deep convolutional neural network} to learn high-level abstract features from CPAP flow signals and a \emph{recurrent neural network} to encode temporal context in these features. The entire \textbf{Neural CRF} (CNN-RNN-CRF) network is trained for sleep staging in an end-to-end fashion.

%\emph{we explicitly model the input signal and the output sleep stages jointly by using a conditional random field (CRF) combined with R-CNN trained 
%in an end-to-end fashion}. 

Our Neural CRF method shows a substantial improvement over the state-of-art when applied to the CPAP flow signal for sleep staging. Further, we improve the performance using a class distribution cost-sensitive prior to deal with the imbalanced distribution of sleep stages and using a domain dependent regularization over the CRF parameters. In summary, we make the following contributions:
\begin{enumerate} [(a)]

\item While the prior deep learning works have entirely focused on extracting best features from the input signals, we demonstrate that jointly modeling the dynamics of the output sleep stages can substantially increase the performance. We improve over the state-of-the-art conditional adversarial model~\cite{zhao2017learning}  by over 10\% in terms of Cohen's Kappa score. Our approach can be added to the existing deep learning models that are competing in the input space.

\item {We propose to use a CNN architecture along with a recurrent layer to extract high-level features from CPAP flow signals. Our architecture is inspired by the popular ResNet \cite{he2016deep} used in computer vision.} %Researchers who work on CPAP flow signal can take our pre-trained model and get the features to support their downstream tasks.}  

\item We present the first study on automatic sleep staging using the respiratory flow signal. Our model has a direct existing application use case - providing healthcare professionals a way to track the patients undergoing sleep apnea treatment through CPAP devices. By directly linking CPAP flow data to sleep stages, this work has the potential to illustrate an improvement in sleep and create an interest in investigating the cognitive and neuronal benefits of adhering to CPAP therapy.%, earlier studies use EEG, cardio-respiratory, actigraphy, and radio-signals.
%\item Currently, the flow signal from CPAP data is not being used by healthcare professionals actively.  Our model provides health-care providers additional sleep health information by providing sleep stage information and accurate sleep metrics like sleep efficiency; \red{@karan: this point seems a bit repeating the previous point. I could be wrong though.}

\end{enumerate}

We organize the rest of the paper as follows. Section~\ref{sec:related} places our work in the context of the existing literature. Section~\ref{sec:model} describes our 
proposed solution in detail. Section~\ref{sec:experiments} lays out our experimental settings. We present our results and analysis in Section~\ref{sec:results}. Finally, we conclude in Section~\ref{sec:conc}.

%REM: looking at plateaus
%Light & Deep: Transitions  from exhaling to inhale

%% file: related.tex
\begin{figure}[t!]
\centering
\begin{tikzpicture}[->,>=stealth',shorten >=1pt,auto,node distance=3cm,
                    thick,main node/.style={circle,draw,font=\sffamily\Large\bfseries}, el/.style = {inner sep=2pt, align=left, sloped}]

  \node[main node,fill=red,text=white,draw=none] (1) {W};
  \node[main node,draw=none,fill=gray!50!black,text=yellow] (2) [below left of=1] {R};
  \node[main node,draw=none,fill=gray!50!black,text=yellow] (3) [below right of=2] {D};
  \node[main node, draw=none,fill=gray!50!black,text=yellow] (4) [below right of=1] {L};
  %\node[main node, draw=none,fill=gray!50!black,text=yellow] (5) [below of=4] {N2};

  \path[every node/.style={font=\sffamily\small}]
    (1) edge node [left] {} (4)
        edge [bend right] node[left] {} (2)
        edge [loop above] node {} (1)
        %edge node [bend left]  {0.04} (5)
    (2) edge node [right] {} (1)
        edge [bend right=10]  node[el,below] {} (4)
        edge [loop left] node {} (2)
        %edge [bend right] node {0.01} (5)
    (3) edge [bend left] node {} (1)
        edge [bend right] node[right] {} (4)
         edge [loop below] node {} (3)
    (4) edge node [left] {} (3)
        edge [loop right] node {} (4)
        edge [bend right=10]  node[el,above] {} (2)
        edge [bend right] node[right] {} (1);
\end{tikzpicture}
 \vspace{-1.25em}
\caption{Transition diagram for OSA patients~\cite{kim2009markov} with  non-REM states 1 and 2 
combined as Light (L) sleep state. Four sleep states shown are: (W)ake, (R)EM, (L)ight and (D)eep.}
 \label{fig:transitiondiag}
 \vspace{-2em}
\end{figure}
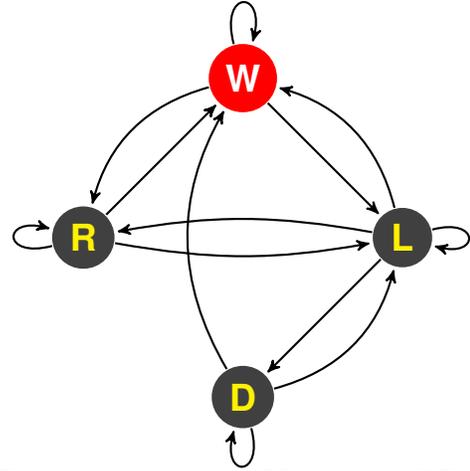

In this section, we describe the different approaches that have been taken towards automating the sleep staging process. PSG is the gold standard for assessment of sleep quality and diagnosis of specific sleep disorders. It requires the subject to spend a night in a sleep lab with a variety of sensors attached to collect data about the biological processes during sleep. %These signals are analyzed in concert to assess the quality and structure of sleep, as well as evaluations for a multitude of sleep-related disorders. 
In clinical practice, several levels of health-care sleep professionals visually annotate the data in 30-second epochs to ascertain the sleep stages. Sleep staging is a labor-intensive process with limitations due to inter-expert variability~\cite{rosenberg2013american}.  There have been some recent efforts to develop automated sleep staging methods.  However, the PSG process has a very high overhead in terms of cost and convenience, and so most of the effort has focused on reducing the number of sensors to asses sleep stages as an alternative to PSG.  We divide the related literature into two parts in the context of our work: (a) signal sources and their application context, and (b) machine learning based sleep staging models.

{\textbf{\Na Signal sources and their application context:}} %We can divide the signal sources into mainly four types: \Ni EEG-based, \Nii Cardio-respiratory sensors, \Niii Actigraphy-based motion sensors, and \Niv No-contact sensors. 
\textbf{EEG-based} sensors measure brain activity and have been shown to indicate sleep states the best, as sleep staging is based on EEG rules during PSGs. %Aboaloayon et. al~\cite{aboalayon2016sleep} present a comprehensive overview of the sleep staging approaches using EEG signal. 
EEG-based sleep staging~\cite{ebrahimi2008automatic, lajnef2015learning,tsinalis2016automatic} has demonstrated the highest accuracy.
%achieving human expert level inter-rater agreements as measured by Cohen's Kappa score. 
However, long-term EEG sensor recording is not practical being both costly and inconvenient. Hence, using more convenient sensors has been proposed.  \textbf{Cardio-respiratory sensors} based methods ~\cite{tataraidze2016sleep,long2014measuring} utilizing the electrocardiogram (EKG) or cardiac impulse signals%, or combinations of both during sleep, for sleep staging. 
have shown moderate accuracy. \textbf{Actigraphy-based} methods~\cite{hedner2004novel,mantua2016reliability} to measure movement 
are good discriminators between wake and sleep but a
poor predictor of sleep stages~\cite{montgomery2012movement}. \textbf{No-contact} sensors like smartphone and radio-signal are the most convenient for the user. Smartphone-based 
approaches~\cite{gu2014intelligent,gautam2015smartphone} have shown to perform poorly like actigraphy.% they can discern non-awake sleep and awake states. 
RF-based approaches have historically demonstrated low accuracy, though recent work by Zhao et al.~\cite{zhao2017learning} demonstrated significant improvements. 

In our work, we use a \textit{new source signal,} namely \emph{\textbf{nasal airflow (flow)}}.  This flow signal is a measure of respiratory effort, similar to chest-band based sensors that measure the respiratory patterns of subjects during sleep. Unlike other methods, however, \textit{the flow signal has an existing use-case}; persons with OSA who are regularly using CPAP therapy. Our method can provide a mechanism for continuously monitoring these patients' sleep health and response to the therapy with significantly improved accuracy for sleep staging and very high accuracy for sleep efficiency. On the clinical research side, we expect that our study motivates researchers for investigating the brain and cognitive effects of CPAP therapy.

{\textbf{\Nb Machine learning based sleep staging models:}} Most of the prior studies using machine learning have used handcrafted features~\cite{lajnef2015learning,tataraidze2016sleep,long2014measuring}. Features based on frequency domain like power spectral density and time domain like variance, skew, or kurtosis have been used mostly as input to classifiers. Recently, \textbf{deep learning} methods using deep convolution 
neural networks~\cite{tsinalis2016automatic,tsinalis2016deep,mikkelsen2018personalizing,chambon2017deep} have been proposed. Hybrid  R-CNN models with CNN as the base network, followed by a recurrent neural network (RNN) have shown state-of-the-art results ~\cite{biswal2017sleepnet,supratak2017deepsleepnet} comparable to expert level annotations on PSG data. Zhao et al.~\cite{zhao2017learning} have proposed an adversarial R-CNN architecture that achieves state-of-the-art results using the radio frequency (RF) signals. 

These deep learning methods have entirely focused on extracting the best possible features from the input signal ignoring or paying limited attention to the context of each segment and dynamics of the sleep states. However, sleep stage transitions have a strong dependency structure~\cite{kim2009markov} with many transitions having extremely low probability. For example, the transition between REM and deep sleep requires an intermediate step; having contiguous REM and deep sleep epochs would require the unlikely occurrence of several transitions taking place inside the same epoch.  Furthermore, some detectable events like rapid eye movements, arousals or K-spindles~\cite{de2003sleep} dictate epoch-to-epoch stability or stage transitions.  In such scenarios, it is essential to take into account both the input signal and the dynamics of sleep states. R-CNN models assume that recurrent connections can capture the sleep state transition structure while attempts with convolution network \cite{chambon2017deep} assume that taking the immediate neighboring segments into account should suffice. 

In this work, we use a \textbf{conditional random field} model that does joint modeling of the sleep stages for the entire duration of sleep, trained end-to-end with a deep R-CNN network. We show that using this approach we can substantially increase the model's performance over adversarial~\cite{zhao2017learning} or baseline R-CNN. Our approach can be augmented to any of the existing deep learning methods. 

%% file: model.tex
In this section, we present our problem and the proposed solution that combines a convolutional neural network (CNN), a recurrent neural network (RNN), and a conditional random field (CRF) in a single architecture that is trained end-to-end. In the following, we describe the components.

\subsection{Problem Statement}
Let the input flow signal time-series be $\mathbf{x}=(x_1,x_2,\dots,x_n)$, 
where $x_i$'s are signal values sampled at 32 Hertz, \ie 32 signal values per 
second. Annotation of sleep stages is done for each 30-second epoch corresponding to $30 \times 32 = 960$ signal values in $\mathbf{x}$. An example flow signal and 
corresponding sleep stages for a night's sleep is shown in Figure~\ref{fig:signalex}. The computational task is to annotate the signal time-series for each 30-second epoch with a label $y_i$ as one of the four sleep stages: \textbf{W}ake, \textbf{R}EM, \textbf{L}ight Sleep, and \textbf{D}eep Sleep. In other words,  we need to label the sequence $\mathbf{x}=(x_1,x_2,\dots,x_n)$ with $\mathbf{y}=(y_1,y_2,\dots,y_m)$,
where $m=n/960$. In our experiments, $m=900$ and $n=900 \times 960 = 864,000$. We denote the set of sleep states as ${K}=\{W,R,L,D\}$, in the rest of the paper.

\subsection{Convolutional Neural Network}
\label{sec:cnn}

%\red{We need a motivation for using CNN. Why is it working? }

Convolution neural networks have been used extensively for a variety of tasks in computer vision, natural language processing, and time-series analysis. %Vanilla convolution operations extract the local features from the input that are relevant to the task at hand. 
We use convolution layers to extract high-level abstract features that are then fed to the recurrent layer. Our convolutional neural network takes the flow signal time-series $\mathbf{x}=(x_1,x_2,\dots,x_n)$ as input and passes it through a sequence of \textbf{convolution layers} to generate $m$ abstract feature vectors $Z = (\mathbf{z}_1, \mathbf{z}_2, \cdots, \mathbf{z}_m)$ that are fed to a recurrent neural network (described in the next subsection) for further processing. 

\begin{figure}[t!]
\centering
\includegraphics[width=0.3\textwidth]{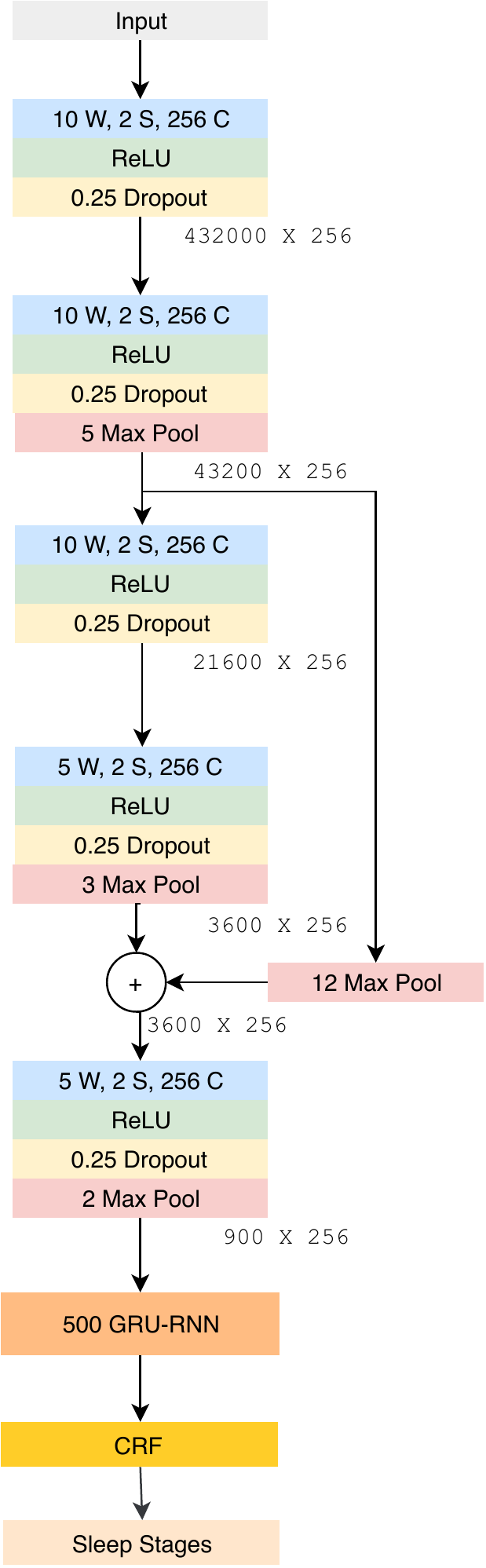}
\vspace{-0.4em}
\caption{Our deep learning model architecture, with CNN layers, GRU connections, and CRF for end-to-end learning. W refers to the Kernel size, S refers to the stride size, and C refers to the channel size at each convolution layer. The numbers indicated at the end of each convolution layer represent the size of output at each layer.}
\label{fig:dnn}
\vspace{-2.5em}
\end{figure}

We use a variation of ResNet architecture~\cite{he2016deep} as our base convolution neural network. Figure \ref{fig:dnn} shows the network. Each {convolution layer} involves a \textbf{1D convolution} operation followed by a rectified linear unit (ReLU) non-linear activation~\cite{nair2010rectified}, a dropout, and (optionally) a max-pooling operation. Let $\mathcal{X} \in \mathbb{R}^{T}$ be the input sequence of length $m$ to a convolution layer (at any depth) with $j$-th kernel $K^j$ of size $W$ and stride size $s$. The $1D$ convolution operation at $t \in \{1, 2, \cdots, T\}$ is defined as:
\begin{eqnarray}\label{eq:cnn}
\vspace{-0.5em}
\Phi_{t}  &=& f( \sum_{i=1}^{W} K_i^j\mathcal{X}_{t+i.s-1}  + b) 
\vspace{-0.5em}
\end{eqnarray}
\noindent where $b$ is the kernel $K$'s bias and $f(\cdot)$ represents the activation function ReLU defined as $f(z)=max(z,0)$. 
The outputs of convolution at each $t$ are concatenated to produce a feature map for each of the $N$ kernels.%, we get $N$ feature maps represented as $\Phi \in \mathbb{R}^{NXO}$, where $O= (T-W)/s+1$. 

Convolution operations help extract the local features of time-series signal in a location invariant way. Strides $s$ and filter width $W$ capture the transitions in the input and receptive field or catchment of the convolution operation, respectively. We  utilize \textbf{dropout} on the rectified activations to avoid over-fitting~\cite{srivastava2014dropout}. For some layers, the convolution-pooling operations are succeeded by a \textbf{max-pooling} operation.% defined as:

%\begin{eqnarray}\label{eq:pool}
%M(\Phi^{j}) &=& [max(\Phi_{1:p}^j),\cdots,max(\Phi_{O-p:O}^j)] \\
%\mathcal{X}' &=& [M(\Phi^{1}),\cdots,M(\Phi^{N})]
%\end{eqnarray}

%\noindent where the $max$ operation selects the maximum value in the $p$-sized vector passed to it, and $\mathcal{X}'$ is the output of the convolution layer,  fed as input to the next layer. The max-pooling selects the most important local features and can also have the option of strides that help control the receptive field of the pooling operation. 

Additionally, we use \textbf{residual connections} between two layers so that a new layer added to the network learns something new. They also help with the diminishing gradient of preceding layers problem in deep convolution networks by forcing the network to learn the identity mapping~\cite{he2016identity}. Formally, let $\mathcal{X}$ be the input to a convolution layer, and  $\mathcal{F(\mathcal{X})}$  represent the output of repeated convolution-pooling layers. The residual connection is defined as:
\begin{eqnarray}\label{eq:pool}
\mathcal{X}'' = \mathcal{F(\mathcal{X})} + U^{\mathbf{T}} \mathcal{X}
\end{eqnarray}
\noindent where $U$ is a transformation matrix that is used to bring $\mathcal{X}$ to the same dimensions as of  $\mathcal{F(\mathcal{X})}$. In case both have the same dimensions, $U$ becomes an identity matrix. Residual connections are usually used after one or two convolution-pooling layers. We use it between second and fourth layers in our network (Figure \ref{fig:dnn}). 

As shown in Figure \ref{fig:dnn}, the first convolution layer in our network uses 256 different kernels (\ie\ output channels) each of size 10 and stride $2$. This generates 432,000 feature values in each feature map. The second convolution layer then applies a kernel of window size 10 and stride 2 to each feature map and reduces the number of features in each feature map to 43,200. This process continues until the last convolution layer, which generates $m=$ 900 features in each feature map. In other words, the output of the CNN is ${Z} \in \mathbb{R}^{256X900}$. 

\subsection{Recurrent Neural Network}
\label{sec:rnn}
{Recurrent neural networks are used to model inputs with sequential nature. Since our data is a time-series sequence, we use the recurrent layer to model the temporal nature of our signal that builds on the high-level features from the convolution layers.}
The \textbf{recurrent} layer takes the feature vectors $Z = (\mathbf{z}_1, \mathbf{z}_2, \cdots, \mathbf{z}_m)$ produced by the preceding ResNet CNN as input, and computes a representation ${\mathbf{h}}_t$ at every time step $t$ by combining the current input  $\mathbf{z}_t$ with the output of the previous time step ${\mathbf{h}}_{t-1}$. The recurrent units model the temporal dynamics of the input signal, by working on the  sharp feature maps of the CNN. 

We use Gated Recurrent Units (GRUs)~\cite{cho2014properties} as our recurrent units.  GRU has two gates: update gate ($u$) and reset gate ($r$) apart from the hidden cell state $h$. It combines the forget gate and the input gate of the popular Long Short-Term Memory (LSTM)~\cite{hochreiter1997long} unit into one update gate. The update equations for GRU can be written as:
\vspace{-0.25em}
\begin{eqnarray}\label{eq:pool}
\mathbf{u}_t &=& \sigma(W_z \mathbf{z}_t + U_z \mathbf{h}_{t-1} + b_z) \\
\mathbf{r}_t &=& \sigma(W_r \mathbf{z}_t + U_r \mathbf{h}_{t-1} + b_r) \\
\widetilde{\mathbf{h}}_t &=& 2\sigma(W_h \mathbf{z}_t + \mathbf{r}_t \odot U_h \mathbf{h}_{t-1} + b_h)-1\\
\mathbf{h}_t &=& \mathbf{u}_t\odot \mathbf{h}_{t-1} + (1-\mathbf{u}_t) \odot \widetilde{\mathbf{h}}_{t}
\end{eqnarray}
where, $\sigma(\cdot)$ is the sigmoid activation function defined as $\sigma(x)=1/(1+e^{-x})$, $W$'s and $U$'s are weight matrices, $b$'s are biases, and $\odot$ denotes the Hadamard or element-wise product.  GRUs have been shown to be much faster owing to reduced number of parameters and perform at par with LSTMs~\cite{jozefowicz2015empirical,chung2014empirical}.%, especially with smaller training data. 

The vector $\mathbf{h}_t$
effectively represents each 30-second epoch in context, which can be used to classify the epoch into one of the sleep stages using a softmax layer. Formally, the probability of $k$-th class for classifying into $K$ {sleep stages} is
\begin{equation}
p(y_t = k|\mathbf{h}_t, W_o) = \frac{\exp~(W_{o,k}\mathbf{h}_t + b_k)} {\sum_{k=1}^{K} \exp~({W_{o,k}\mathbf{h}_t} + b_k)} \label{eq:softmax}
\end{equation}
\noindent  where $W$ are the classifier weights, and $b$ are the bias terms. We minimize the negative log likelihood (NLL) of the gold labels. The NLL for one data point $(\mathbf{x}, \mathbf{y})$ is:
\begin{eqnarray}
\Ls_c(\theta) = - \sum_{k=1}^{K} \sum_{t=1}^{m} \Is(y_t = k) \log p(y_t = k|\mathbf{x}, \theta) \label{eq:logloss}
\end{eqnarray}

\noindent where $\theta$ denotes the set of model parameters, and $\Is(y = k)$ is an indicator function to encode the gold labels: $\Is(y = k)=1$ if the gold label $y=k$, otherwise $0$.\footnote{This is also known as one-hot vector representation.} The loss function minimizes the cross-entropy between the predicted distribution and the target distribution (\ie\ gold labels). We refer to this combined architecture (\ie\ an RNN layer on top of a CNN) as \textbf{R-CNN}.

%In our model, we feed the output of each 30-second epoch from CNN into our RNN for prediction of sleep stage for each 30-second epoch segment using a softmax layer. 

\subsection{Conditional Random Field}

%In the case of conditional random field (CRF) network, the hidden state of each GRU cell is fed into the CRF, without a preceding softmax layer prediction. Next, we describe CRF formulation.

%\red{Should start with a motivation. Also the limitations of R-CNN}

\subsubsection{\textbf{Motivation}}
 The R-CNN model presented above works in the input signal space and predicts the sleep stage for each time step independently based on the corresponding RNN hidden state. Although it considers the input context through recurrent layers, it is oblivious to the dynamics in 
the output space, \ie dynamics of the sleep stages. 

Prior works using deep neural networks have focused entirely on extracting the best features from 
the input signals like EOG, ECG, or RF signals for predicting the sleep stage~\cite{biswal2017sleepnet,supratak2017deepsleepnet,zhao2017learning,tsinalis2016automatic,tsinalis2016deep,mikkelsen2018personalizing}. Like R-CNN, these methods make independent (as opposed to collective) decisions. We argue that this approach is not optimal especially when there are strong dependencies across output labels. It is known that the sleep stage transitions have a strong dependency structure~\cite{kim2009markov}. {For example, a number of transitions are not allowed as can be seen in Figure~\ref{fig:transitiondiag}. Also, there could be complex dependencies like the long-term cyclical effect of events like arousal or K-complex spindles on deep and REM sleep states~\cite{de2003sleep}}. Exploiting this transition structure for an accurate sleep staging is important. Also, because of local normalization (\ie\ softmax in Equation \ref{eq:softmax}), these models suffer from the so-called label bias problem \cite{lafferty2001conditional}.     

Instead of modeling classification decisions
independently, we model them jointly using a conditional random field or \textbf{CRF}~\cite{lafferty2001conditional}. In our network, we put the CRF layer above the recurrent layer of R-CNN, and train the whole network end-to-end. CRFs have been shown to be able to utilize the global temporal context for maximizing 
sequence probabilities, relying upon first- or higher-order Markov assumptions over the output label transitions. 

\subsubsection{\textbf{Neural CRF}}
The input to our CRF layer is a sequence of hidden states $H = (\mathbf{h}_1, \mathbf{h}_2, \cdots, \mathbf{h}_m)$ from the GRU-based recurrent layer, and the corresponding label sequence is $\mathbf{y} = ({y}_1, {y}_2, \cdots, {y}_m)$. The compatibility of the input feature $H$ and an output label $y_t \in \{W, R, D, L\}$ at time step $t$ is computed by the unary (node) potential defined as:
\begin{eqnarray}\label{eq:pool}
\mathbf{\Psi}_n(y_t|H,\mathbf{w}_n,b_n) = \mathsf{exp}({\mathbf{w}_n^{T} \phi(y_t,H)}+b_n)
\end{eqnarray}
\noindent where $\phi(\cdot)$ denotes the feature vector computed from the input and the sleep stage labels, and $\mathbf{w}_n$ is the associated weight vector. Here, $\mathbf{\Psi}_n(y_t|H,\mathbf{w}_n,b_n)$ can be considered as a score (unnormalized probability) given to label $y_t$. Applying the node potential to all nodes in the sequence generates a matrix $S$ of size $m \times |K|$, where $K$ is the set of sleep stages/classes (in our case $|K|=4$), and $S_{i,j}$ corresponds to the score of the $j$-th class for input $\mathbf{h}_i$.  

To model dynamics in the label sequence, we define edge potentials between $y_{t-1}$ and $y_{t}$  as:
\begin{eqnarray}\label{eq:pool}
\mathbf{\Psi}_e(y_{t-1}, y_t|H,\mathbf{w}_e,b_e) = \mathsf{exp}({\mathbf{w}_e^{T} \phi(y_{t-1},y_t,H)}+b_e)
\end{eqnarray}
\noindent where $\phi(y_{t-1},y_t,H)$
denotes edge features with $\mathbf{w}_e$ being the corresponding weight vector. The edge potential computes a score for each possible edge transition in a matrix of size $|K| \times |K|$. The joint conditional probability for the sequence is defined as:
\vspace{-0.25em}
\begin{eqnarray}\label{eq:crfprob}
p(\mathbf{y}|H,\theta) & = & \frac{1}{Z(H,\theta)}\prod_{t=1}^{m} 
\mathbf{\Psi}_n(y_t|H,\mathbf{w}_n,b_n) \nonumber \\ 
& & \prod_{t=2}^{m} \mathbf{\Psi}_e(y_{t-1},y_t|H,\mathbf{w}_e,b_e) 
\end{eqnarray}
\noindent where $Z(H,\mathbf{w}_n,\theta)$ is the global normalization constant (partition function) derived as the sum over all  possible sequences and $\theta$ denotes the set of all parameters in the complete (R-CNN-CRF) network.
\vspace{-0.25em}
\begin{eqnarray}
 {Z(H,\theta)} = \sum_{\mathbf{y}} \prod_{t=1}^{m} 
\mathbf{\Psi}_n(y_t|H,\mathbf{w}_n,b_n)  \nonumber \\  
 \hspace{4em} \prod_{t=2}^{m} \mathbf{\Psi}_e(y_{t-1},y_t|H,\mathbf{w}_e,b_e)
\end{eqnarray}
This global normalization constrains the distribution to a valid probability distribution and helps overcome the label bias problem of locally normalized models. The negative log-likelihood for one data point can be expressed as:
\vspace{-0.25em}
\begin{eqnarray}
\Ls(\theta)  =  - \log   p(\mathbf{y}|H,\theta) \\
= \mathrm{log~Z} -\sum_{t=1}^{m}\mathbf{w}_n^{T} \phi(y_t,H) - b_n \nonumber \\
- \sum_{t=2}^{m} {\mathbf{w}_e^{T} \phi(y_{t-1},y_t,H) - b_e} \label{eq:crfloss}
\end{eqnarray}
\noindent Note that the objective in Equation \ref{eq:crfloss} is convex with respect to the CRF parameters $\theta' = \{\mathbf{w}_e,\mathbf{w}_n,b_e,b_n\}$ assuming the 
inputs from the R-CNN (\ie\ $Z$) are fixed. In training, we add a $l_1$ regularization on the CRF parameters $\theta'$  to promote sparsity. The final objective can thus be written as:
\vspace{-0.25em}
\begin{eqnarray}\label{eq:crfnll}
\min_{\theta} \Ls(\theta) + \lambda \norm{\theta'}_1
\end{eqnarray}
As can be seen in 
Figure~\ref{fig:transitiondiag}, a number of transitions are not observed while some have a large value making $l_1$ norm more appropriate to our problem compared to a $l_2$ norm. 

The complete R-CNN-CRF network is trained end-to-end on the loss in Equation~\ref{eq:crfnll} by back-propagating the errors (gradients) to the R-CNN. Similar end-to-end training has shown impressive results in computer 
vision~\cite{zheng2015conditional}. Once the parameters of the network are learned, decoding the most probable sequence is performed effectively using the Viterbi algorithm.

\subsection{Class Distribution Cost-Sensitive Prior}
After analyzing our training data, we found that the class distribution of different sleep stages is skewed with REM and Deep sleep forming less than 10\% of annotations each. To tackle this issue, we add a class prior $\alpha_k$ over the log likelihood. The class prior $\alpha_k$ for $k \in \{1, 2, \ldots, K\}$ is estimated from the training data by: 
\vspace{-0.25em}
\begin{eqnarray}\label{eq:prior}
\alpha_k = \frac{n_{\mu}}{n_k}
\end{eqnarray}
\noindent where $n_{\mu}$ is the average number of labels in each class \ie $n/K$ with $n$
being the total number of sleep labels in the training set. We incorporate the class prior in our loss from Equation \ref{eq:crfnll} as: 
\vspace{-0.25em}
\begin{eqnarray}\label{eq:crfnllprior}
\min_{\theta} ~ - \sum_{k=1}^K  \sum_{t=1}^m \Is(y_t = k) \alpha_k \log~p({y_t} = k|\theta)  + \lambda \norm{\theta'}_1
\end{eqnarray}
\noindent where $\Is(.)$ is the boolean indicator function as defined before.  
The priors $\alpha$ being inversely proportional to the number of samples of the class giving more weight-age to the under-represented classes leading to balanced learning during the training phase. We demonstrate the benefit of this prior in our experiments. 

%% file: experiments.tex
In this section, we describe our dataset, the metrics used, and baseline methods we compare our approach against.

\subsection{Dataset}
We use the publicly available data from Multi-Ethnic Study of Atherosclerosis (MESA)~\cite{bild2002multi} provided by National Sleep Research Resource (NSRR) as part of an
initiative to provide the health informatics community for building tools that can be helpful for sleep research and health-care~\cite{dean2016scaling}. We use the nasal airflow pressure channel 
data (univariate time-series), hereto referred to as flow signal, from PSG data of 400 randomly sampled sleep apnea patients. 

\begin{figure}[t]
\centering
\includegraphics[width=0.45\textwidth]{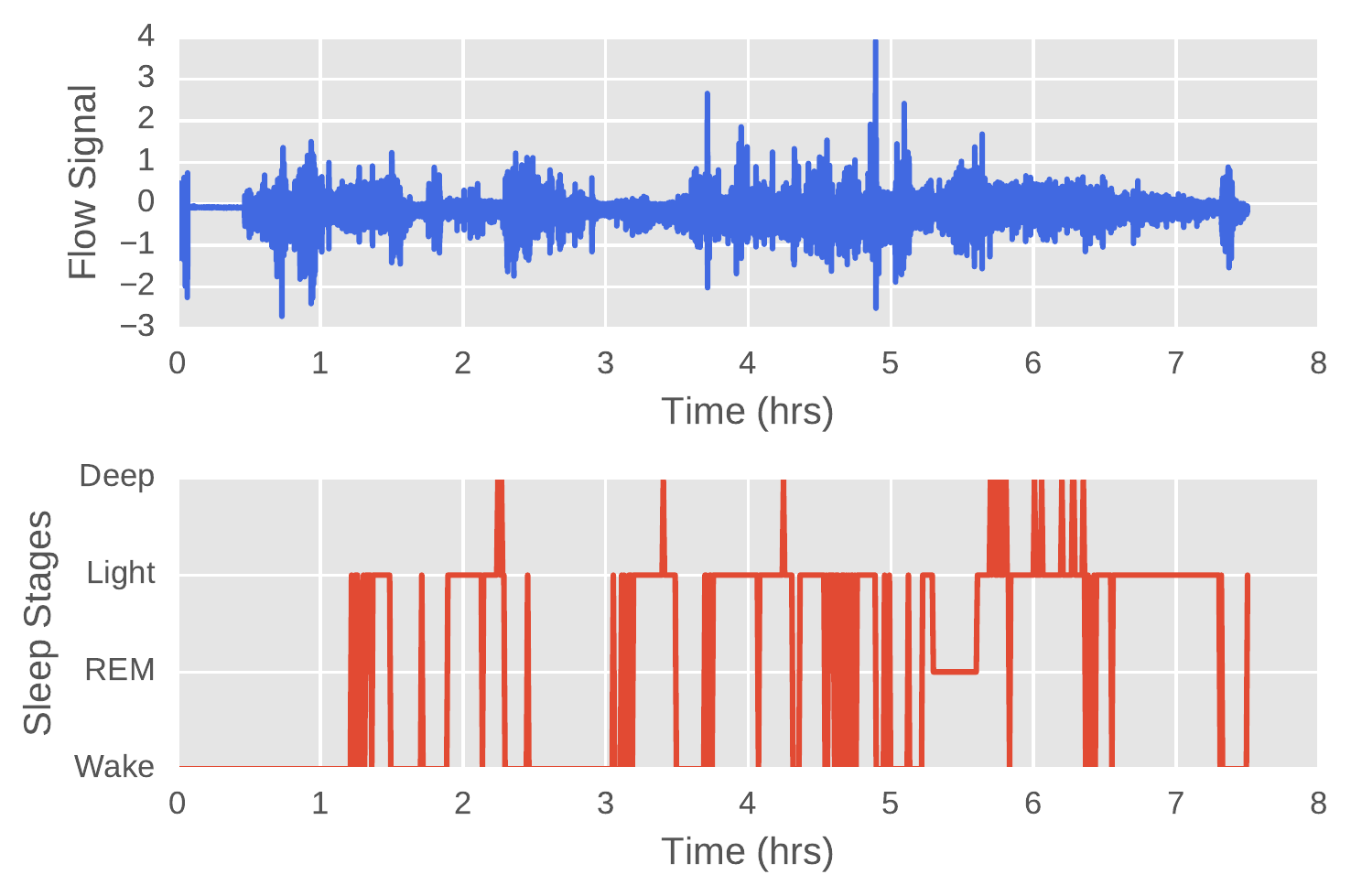}
\caption{An illustrative example of flow signal and corresponding sleep stage annotations for a subject.}
\label{fig:signalex}
\vspace{-2em}
\end{figure}

The flow signal is 
sampled at 32 Hz (\ie 32 samples per second), while the sleep stages are annotated for contiguous 30-second epochs during the duration of the recording. The dataset   
comes with five sleep stages annotated: wake, REM, N1, N2, and N3/4. At the time of the MESA cohort collection, stages N3 and N4 were distinct entities.  However, subsequent research demonstrated lack of fundamental differences between N3 and N4, and N4 has been subsumed into N3, otherwise known as deep sleep~\cite{silber2007visual}. Prior studies have generally combined stages N1 and N2 into a unified group, light sleep, due to transitory nature of N1 and difficulty in differentiating between the two states. We follow the same approach, and we identify four states: wake, REM, light, and deep sleep. We take 7.5 hours of flow data for each patient for our experiments. 
An example of a single subject data including flow signal and sleep staging is shown in Figure~\ref{fig:signalex}.

\subsection{Baseline Methods}
We use the vanilla linear-chain CRF\cite{lafferty2001conditional} and the R-CNN models as our baseline methods to compare against our model.

\begin{enumerate}[(a)]

\item \textbf{Conditional Random Field (CRF)}: We provide power spectral density features with the high-frequency band of 16 Hz as input to the CRF. We train it with Limited-memory BFGS learning algorithm with elastic net regularization. We use this baseline to show the benefit of using the deep network as a feature extractor instead of using the handcrafted features.

\item \textbf{R-CNN}: It has the same architecture as described in Subsections \ref{sec:cnn} - \ref{sec:rnn}, with a softmax as the output layer at each time step. R-CNN  classifies each time step independently and serves as the baseline deep learning method.

\item \textbf{Conditional Adversarial Resnet-LSTM}: We implemented and trained 
the conditional adversarial R-CNN (ResNet with LSTM-RNN) 
architecture~\cite{zhao2017learning} until the discriminator loss stabilizes 
to entropy. Their network was shown to be able to remove 
the environmental noise from the representations learned by the network with 
state-of-the-art results using \emph{Radiofrequency (RF)} based signals.

\item \textbf{R-CNN with attention:} We added attention mechanism on top of our base R-CNN model. We used soft dynamic attention {~\cite{MeiBW16}} in our model with a local window.

% instead of attending at the global level because of the nature of our sequence. 
%We did not see any improvement in kappa score and hence didn't include it in further experiments.

\end{enumerate}

\subsection{Variants of Our Neural CRF Model}

We experiment with different variants of our neural CRF model.

\begin{enumerate}[(a)]

\item \textbf{Neural CRF}: This is the base R-CNN network augmented with linear-chain CRF, and trained end-to-end according to Eq.~\ref{eq:crfloss}. 

\item \textbf{Second Order Neural CRF}: It builds up on the Neural CRF with second order edges of the form $(y_t,y_{t+2})$ in addition to the first order edges $(y_t,y_{t+1})$, thus captures longer dependencies.

\item \textbf{Cost-sensitive Neural CRF}: It takes into consideration the class distribution priors from the training dataset according to Eq.~\ref{eq:crfnllprior} without using the $l_1$ regularization.%, \ie $\lambda=0$.

\item \textbf{Regularized cost sensitive Neural CRF}: This regularizes the cost-sensitive Neural CRF with $l_1$ regularization (Eq.~\ref{eq:crfnllprior}).

\end{enumerate}

%Next, we describe the evaluation metrics used to judge the performance of our models.

\subsection{Evaluation Metrics}

Cohen's Kappa coefficient ($\kappa$) and accuracy are the two commonly used metrics to compare a sleep staging model's predictions with ground truth annotations from PSG.  In addition, we report the mean absolute error (MAE) of sleep efficiency. % We define these metrics as below:

\begin{enumerate}[(a)]
\item \textbf{Accuracy}: Accuracy is defined as a fraction of correct labels predicted by the model out of total number of annotations. %Formally, accuracy is defined as:
%\begin{eqnarray}\label{eq:acc}
%\mathrm{Accuracy} = \frac{n_{c}}{N}
%\end{eqnarray}

%where, $n_c$ is the number of labels correctly predicted and $N$ is the total number of annotations.

\item \textbf{Kappa}: Cohen's Kappa coefficient ($\kappa$)~\cite{cohen1960coefficient} is a commonly used metric for sleep stage prediction quality that accounts for blind luck in 
model prediction. It measures the degree of concordance between two independent raters - model prediction and ground truth distribution. %It is defined as:
%\begin{eqnarray}\label{eq:kappa}
%\kappa = \frac{p_o - p_c}{1-p_c} = \frac{\mathrm{Accuracy} - p_c}{1-p_c}
%\end{eqnarray}
%where $p_o$ is the observed agreement of the mode with the ground truth, \ie accuracy, while $p_c$ is the probability of chance agreement based on observing 
%any of the sleep stages.
$\kappa$ has values ranging from 0 to 
1, with 0 being agreement by pure luck and 1 being a total convergence between the raters. A $\kappa$ score of $<$0.4 is considered low, $>$0.4 moderate, $>$0.6 high, and $>$0.8 to be 
near perfect agreement with observed data. The sleep staging ground truth annotations by two or more human experts has a $\kappa$ of \textbf{0.85} on our dataset. Thus, we can be very confident of the 
sleep stage annotations provided by MESA. 

\item \textbf{Mean Absolute Error (MAE) of Sleep Efficiency}: Sleep efficiency is a common metric used by sleep researchers for assessing sleep quality. It is 
defined as a fraction of time with non-wake sleep over the total duration of sleep. 
One of the uses of sleep staging is to calculate the sleep efficiency metric as a first pass metric over the quality of sleep. Sleep efficiency is given by:
\begin{eqnarray}\label{eq:kappa}
\mathrm{SE} = \frac{n_R+n_L+n_D}{n_A+n_R+n_L+n_D}
\end{eqnarray}
where, $n_A$,$n_R$,$n_L$,$n_D$ refer to number of 30-second epochs spent in wake, REM, light, and deep sleep, respectively. We calculate the absolute error of sleep efficiency for a patient $p \in \mathcal{P}$ as the absolute difference between estimated sleep efficiency ($\widehat{SE_p}$) from the predicted sleep stage labels and real sleep efficiency ($SE_p$). The MAE over the test set $\mathcal{P}$ is defined as:
\begin{eqnarray}\label{eq:mae}
\mathrm{MAE} = \frac{1}{|\mathcal{P}|} \sum_{p \in \mathcal{P}} \frac{ | \widehat{SE_p} - SE_p |}{SE_p}
\end{eqnarray}
While we do not train our models to optimize MAE, we assess their performance with MAE. Accurate calculation of sleep efficiency is one of the applications of sleep staging that is very useful for both the patient and medical practitioner to monitor progress on CPAP therapy.  

\end{enumerate}

\subsection{Hyper-parameter Tuning}
We split our dataset based on subjects into 60\% for training, 20\% for validation, and 20\% for testing, \ie we never train on data from a subject in the test dataset. We tune the hyperparameters on the validation set by using \emph{early stopping} with parameters that provide the best validation $\kappa$ score. The parameters chosen for the R-CNN model are shown in Figure~\ref{fig:dnn}. We experimented with different choices of CNN layers --- kernel sizes (5, 10, 15, 20), strides (2, 3, 4, 5), channel sizes (128, 256, 512), and residual connections (1-to-3, 2-to-4, {ResNeXt}~\cite{xie2017aggregated}) for our 5-layered CNN.  For RNN, we experimented with  hidden dimensions of 125, 250, 300, 500, and 750. We chose the regularization coefficients $\lambda$ to be 0.005 after tuning on $\lambda \in \{0.01,0.005, 0.0001\}$. Our implementation uses TensorFlow~\cite{abadi2016tensorflow}. %Next, we present our results based on just described settings.

\begin{figure*}[b!]
\centering
\begin{subfigure}{.3\textwidth}
\centering
\includegraphics[width=.8\linewidth]{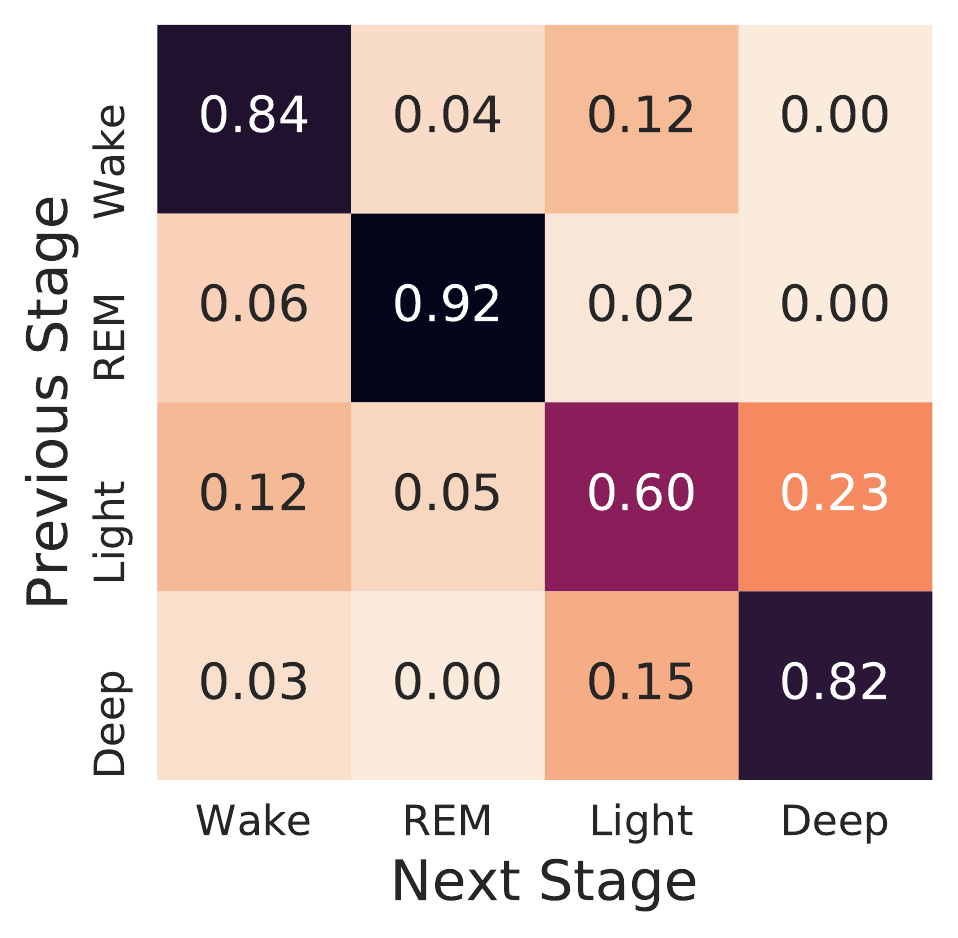}
\caption{CRF sleep stage transition matrix}
\label{fig:trans_crf}
\end{subfigure}
\begin{subfigure}{.3\textwidth}
  \centering
  \includegraphics[width=.8\linewidth]{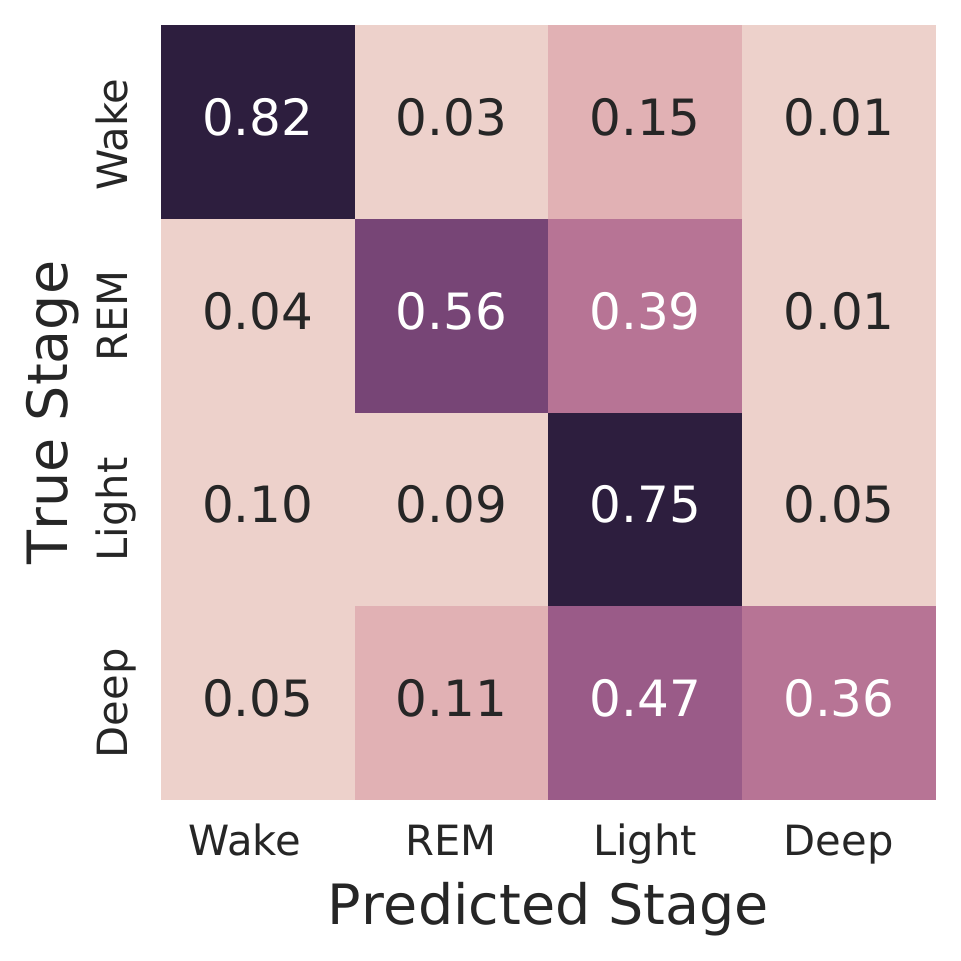}
  \caption{ Neural CRF model}
  \label{fig:conf}
\end{subfigure}%
\begin{subfigure}{.3\textwidth}
  \centering
  \includegraphics[width=.8\linewidth]{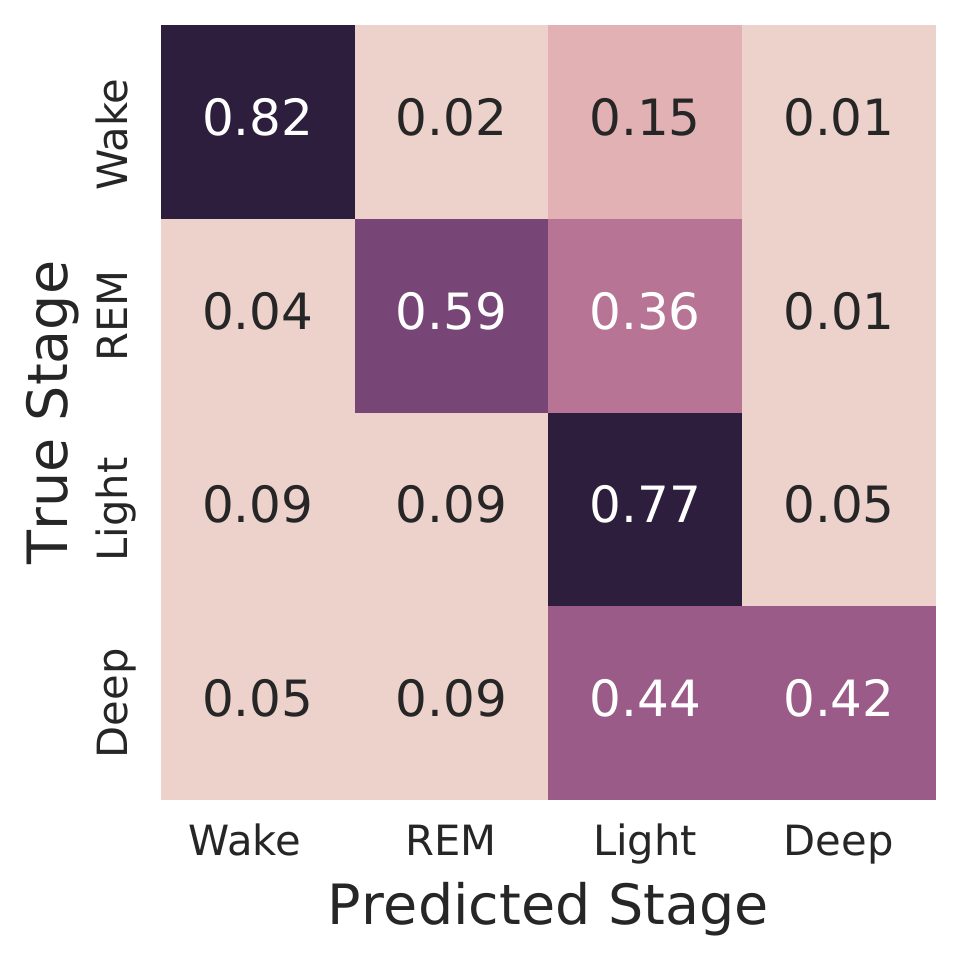}
  \caption{Cost-Sensitive Neural CRF model }
  \label{fig:conf_cs}
\end{subfigure}
\caption{Transition matrix from CRF of Regularized Cost-Sensitive Neural CRF from training (left). Confusion matrices of prediction from the baseline Neural CRF model (center) and cost-sensitive Neural CRF model (right) on the test dataset. }
\label{fig:conf_matrix}
\end{figure*}

%% file: results.tex
In this section, we present our findings on sleep stage classification and compare with the baselines and existing methods. We also present an analysis of our findings with the effect of cost-sensitive prior and saliency map visualizations of model predictions.

\subsection{Classification and Sleep Efficiency Results}
We categorize our methods into three: \Ni linear (non-deep) CRF, \Nii deep neural models with the softmax output layer, and \Niii our deep neural models with CRF output layer.
The results are shown in Table~\ref{table:results}. We summarize our findings below.

\begin{table}[t!]
\vspace{-1.5em}
\scalebox{0.9}{\begin{tabular}{l|ccc}  
\textbf{Approach} & \multicolumn{1}{c}{\bf{Accuracy}} & $\mathbf{\kappa}$ & \bf{SE MAE\%}\\
\midrule
%\multicolumn{4}{|c|}{Non-deep learning based} \\\hline
CRF & 52.4\% & 0.28 & 29.4\% \\\hline
%\multicolumn{4}{|c|}{Deep learning based} \\\hline
R-CNN & 71.5\% & 0.49 & 12.5\%\\
Conditional Adversarial~\cite{zhao2017learning} & 71.1\% & 0.49 & 12.6\%\\
Attentional R-CNN & 70.7\% & 0.48 & 12.8\% \\
\midrule
Neural CRF & 72.3\% & 0.54 & 10.9\%\\
Neural CRF (Order 2 ) & 72.5\% & 0.55 & 10.8\%\\
Cost Sensitive Neural CRF & 73.9\% & 0.56 & 10.3\%\\
\textit{Regularized} Cost Sensitive Neural CRF & \textbf{74.1\%} & \textbf{0.57} & \bf{9.9}\%\\
\bottomrule
\end{tabular}
}
\caption{Accuracy, Kappa, and Sleep efficiency (SE) MAE (lower the better) scores for different approaches.}
\label{table:results}
\vspace{-2.5em}
\end{table}

%\vspace{0.2em}
\noindent \Ni \textbf{Linear (non-deep) CRF:} As we can observe, the baseline CRF  performs poorly with a low accuracy of 52.4\%, $\kappa$ of only 0.27, and higher MAE of 29.4 on sleep efficiency. It 
can be attributed to the low representational power of the input features compared with the task-specific feature extraction of deep learning architectures. 

%\vspace{0.2em}
\noindent \Nii \textbf{Deep neural models with softmax output:} The deep neural models improve the performance considerably over the non-deep CRF model by taking the accuracy to 71\%, $\kappa$ to 0.49, and MAE down to 12.5. Our baseline R-CNN with ResNet-GRU architecture performs the best overall. The conditional adversarial network ~\cite{zhao2017learning} performs at par with the R-CNN, while it poses additional training challenges because of the instability caused by adversarial training \cite{ArjovskyCB17}. Using the local attentional mechanism~\cite{MeiBW16} leads to a slight drop in performance, possibly due to the extra parameters. The residual connection described in Section~\ref{sec:cnn} was quite beneficial for the R-CNN model; it increased the $\kappa$ scores from 0.29 to 0.49.

%\vspace{0.2em}
\noindent \textbf{Remark:} Our experiments that used LSTM units and bi-directional GRU/LSTM as recurrent units in R-CNN did not make a difference in the performance. To train faster, we use unidirectional GRUs in our experiments.

\vspace{0.2em}
\noindent \Niii \textbf{Neural CRF models:} Adding the CRF layer to the base R-CNN improves the performance substantially   taking the $\kappa$ score to 0.54, a 10.2\% 
increase in relative terms. Adding second order edges in the CRF marginally improves the performance, though it increases 
the run-time of the model substantially. {Using a cost-sensitive version of the Neural CRF increases the performance considerably by 2\% in $\kappa$ over the Neural CRF, while the regularized cost sensitive CRF improves the performance by 4\% over the Neural CRF model. Using CRF that infers the global temporal context improves the performance substantially. Adding domain dependent prior knowledge like cost-sensitive prior and sparse regularization helps bring additional gains in the model performance.}  While the increase in \textit{accuracy may seem marginal}, the increase in \textbf{$\kappa$ of 14\% is substantial} since it reflects an improved detection of the difficult and less frequent deep and REM sleep stages.

Transition matrix of the regularized Neural CRF model is shown in Figure~\ref{fig:trans_crf}. As we can see the values 
in the matrix assign zero scores to the non-existent transitions in the
Figure~\ref{fig:transitiondiag}. We also found that these values are close to the ground truth transition values we observe in the dataset.

%\emph{From the performance of our neural CRF models, we can assert that leveraging the global temporal context of input and output jointly is vital for sequence labeling tasks with strong inherent dynamics in the output space. Adding domain dependent prior knowledge like cost-sensitive prior and sparse regularization helps bring additional gains in the model performance.}  

The deep non-linear layers help the model to extract the feature space that is very relevant to the task as reflected by the difference in performance of the non-deep 
CRF vs. R-CNN. Combining the two as our Neural CRF does is very helpful in leveraging the strengths of both approaches -- deep learning for meaningful feature extraction in the input side, and the modeling strength of CRFs, which use global inference to model consistency in the output structure.

The precisely similar trend as above is observed with sleep efficiency MAE. Our models can predict sleep efficiency metric with a reasonable accuracy --- within 10\%-15\% of the sleep efficiency value. This is expected since the model is able to differentiate the wake state with very high accuracy as shown by the confusion matrices in Figures~\ref{fig:conf} and~\ref{fig:conf_cs}. Hence, 
our model can provide an accurate estimation of sleep efficiency to help health-care professionals track the response of CPAP therapy.

\begin{figure}[t!]
\centering
\vspace{-1em}
\includegraphics[width=0.3\textwidth]{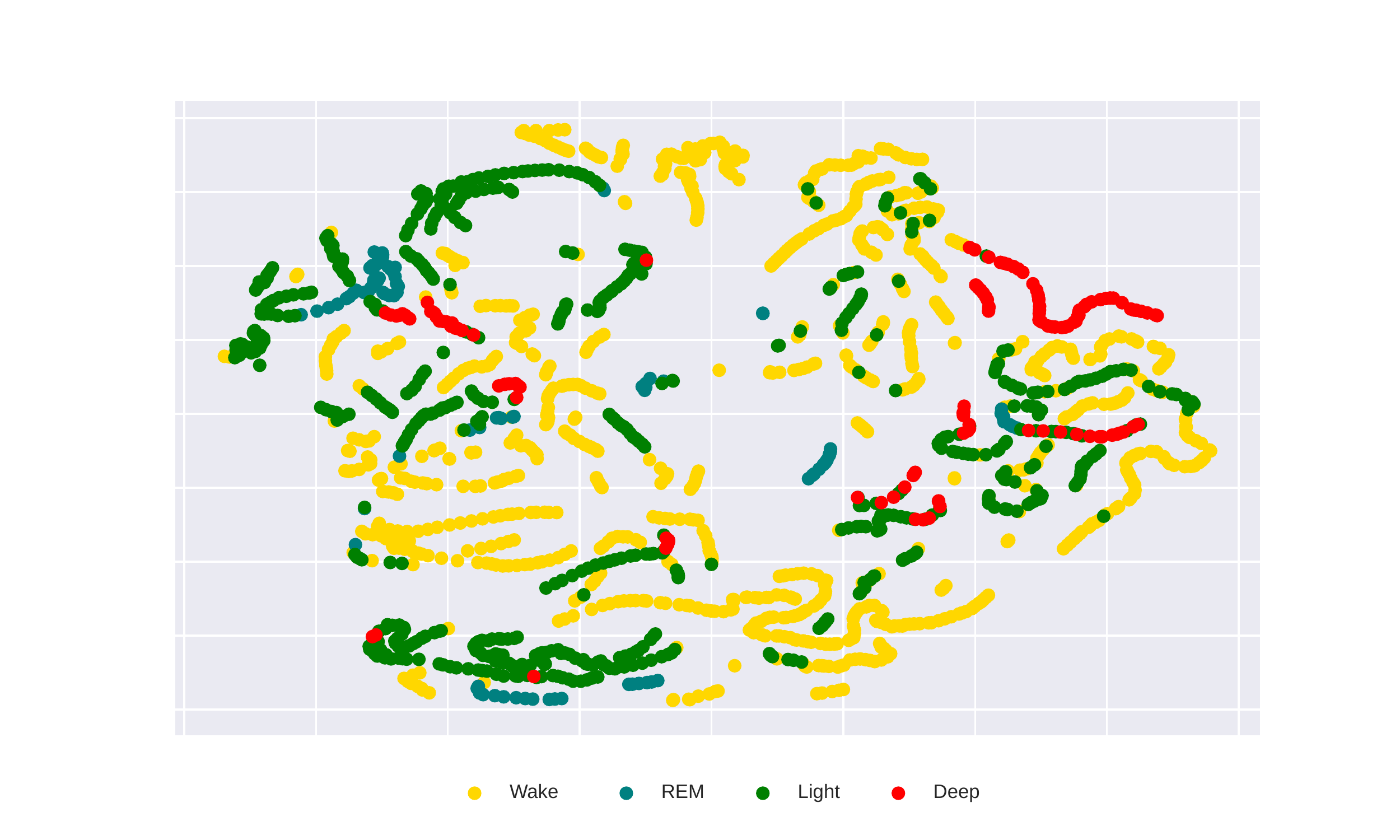}
\caption{t-SNE plot of GRU-RNN output from Regularized Cost-Sensitive Neural CRF for each annotation. RNN is able to differentiate the wake and light states to a good measure.}
\label{fig:tsne}
\vspace{-1.75em}
\end{figure}

%\begin{figure}[t]
%\centering
%\includegraphics[width=0.45\textwidth]{figures/cm_b.pdf}
%\caption{Confusion matrix of prediction from the baseline R-CNN model.}
%\label{fig:conf_matrix}
%\end{figure}
%\begin{mdframed}[style=myframe]
%Leveraging the global temporal context of input and output jointly
%is important for sequence labeling tasks with a strong inherent dynamics in the output space. 
%By capturing the output sleep stages dynamics using conditional random field end-to-end with a competent input feature extractor like deep neural network  improves the sleep staging accuracy substantially.
%\end{mdframed}

\begin{figure*}[t!]
\centering
\begin{subfigure}{.23\textwidth}
  \centering
  \includegraphics[width=\linewidth]{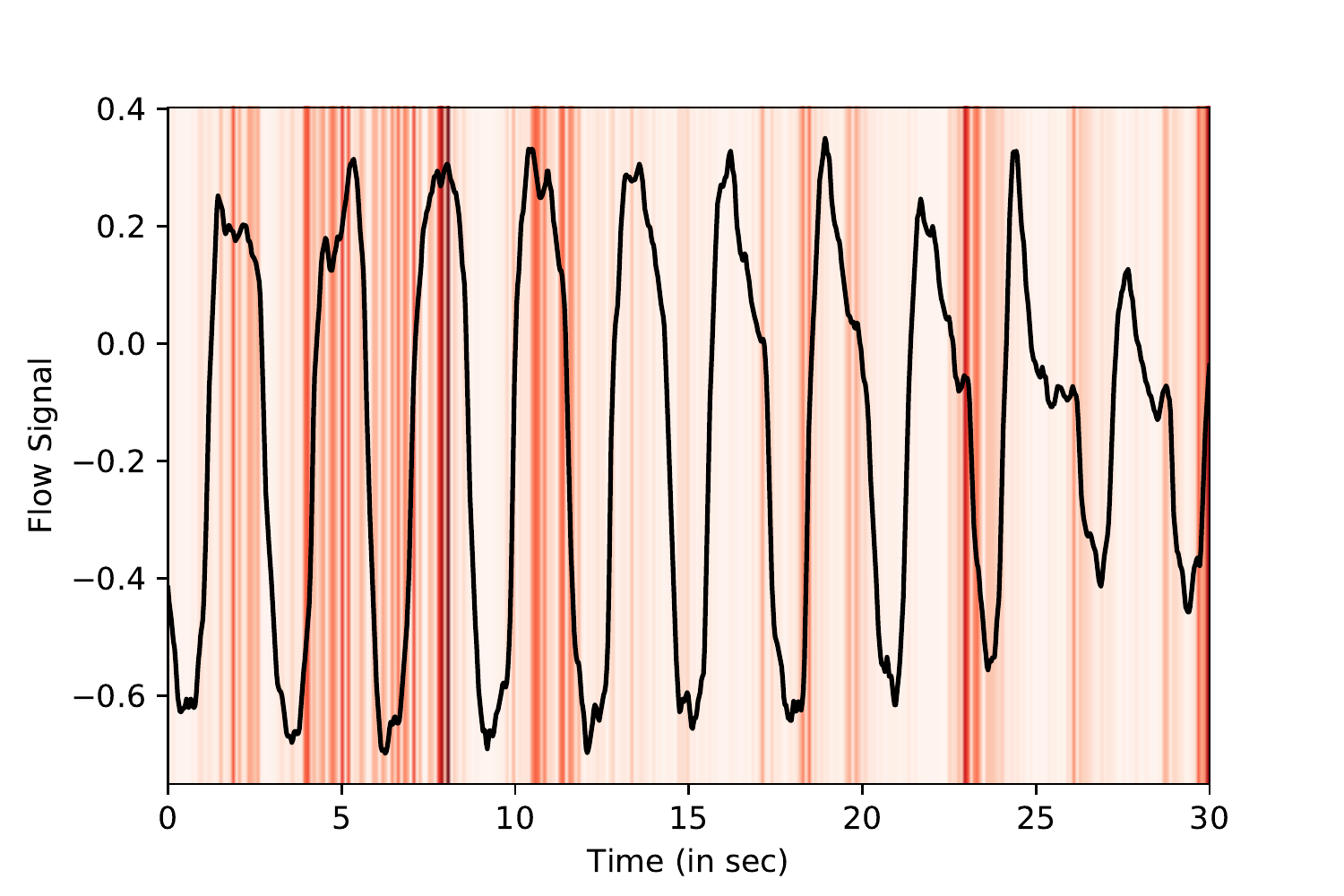}
  \caption{ Wake}
  \label{fig:awake_sal}
\end{subfigure}%
\begin{subfigure}{.23\textwidth}
  \centering
  \includegraphics[width=\linewidth]{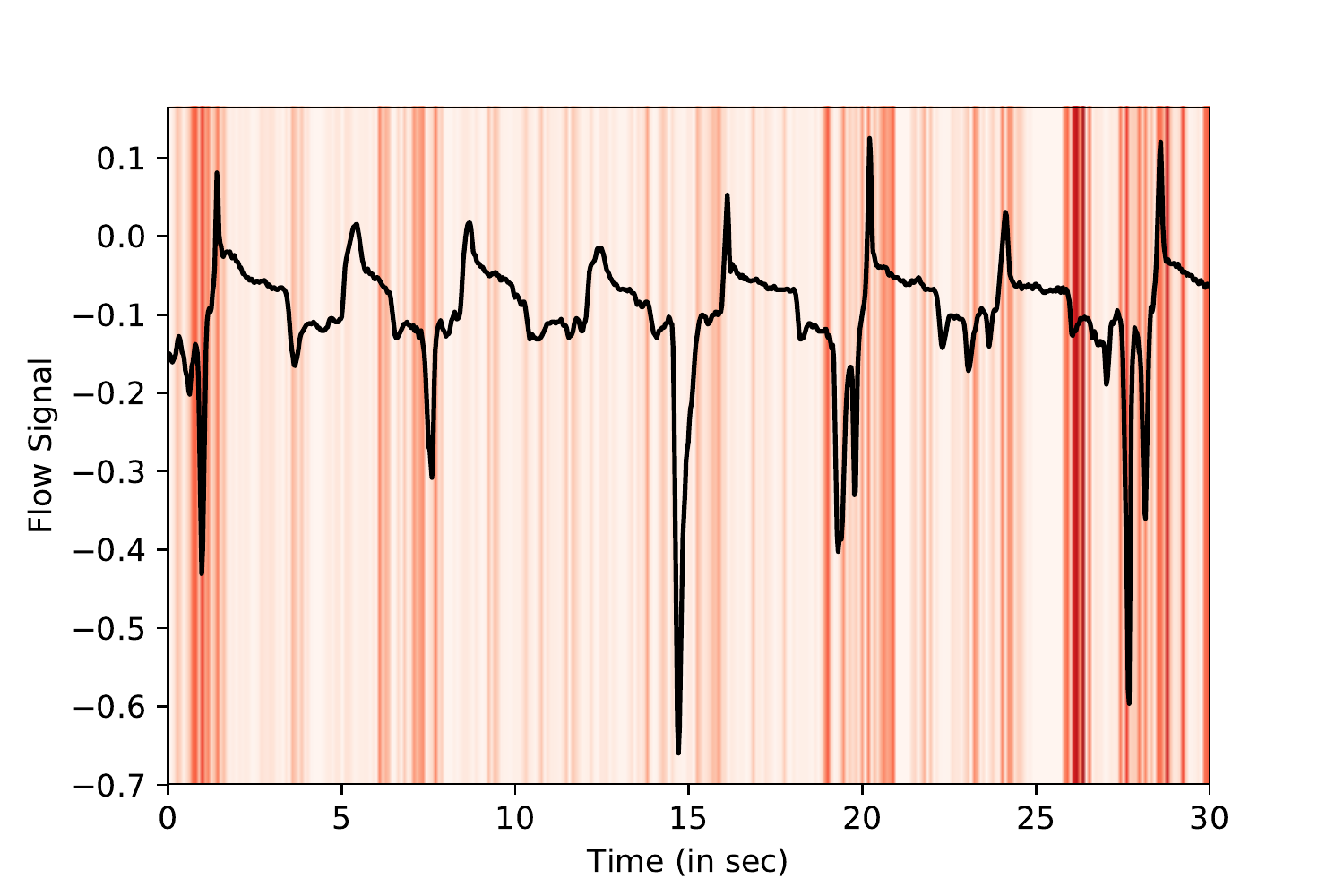}
  \caption{REM }
  \label{fig:rem_sal}
\end{subfigure}
\begin{subfigure}{.23\textwidth}
\centering
\includegraphics[width=\linewidth]{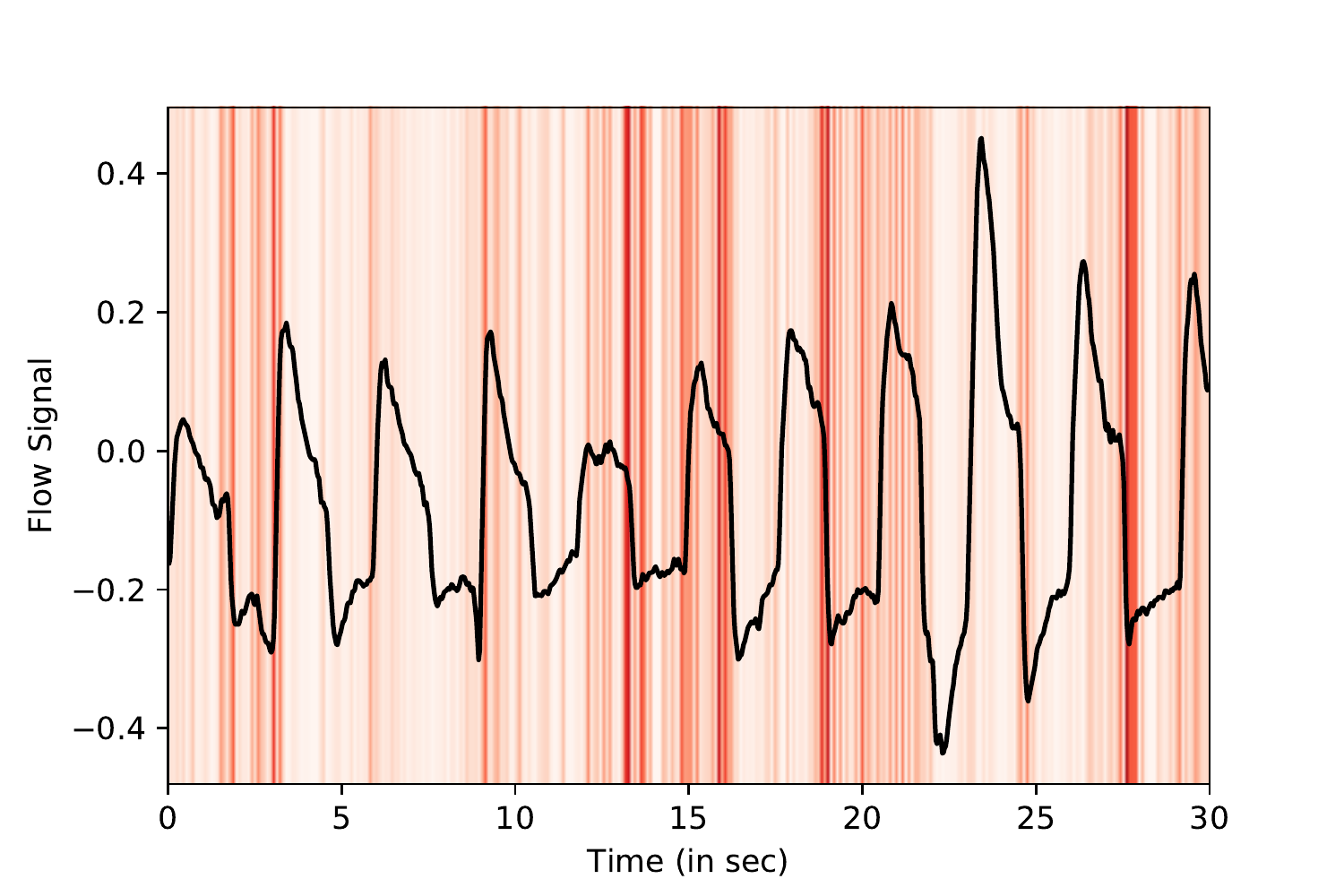}
\caption{Light}
\label{fig:light_sal}
\end{subfigure}
\begin{subfigure}{.23\textwidth}
\centering
\includegraphics[width=\linewidth]{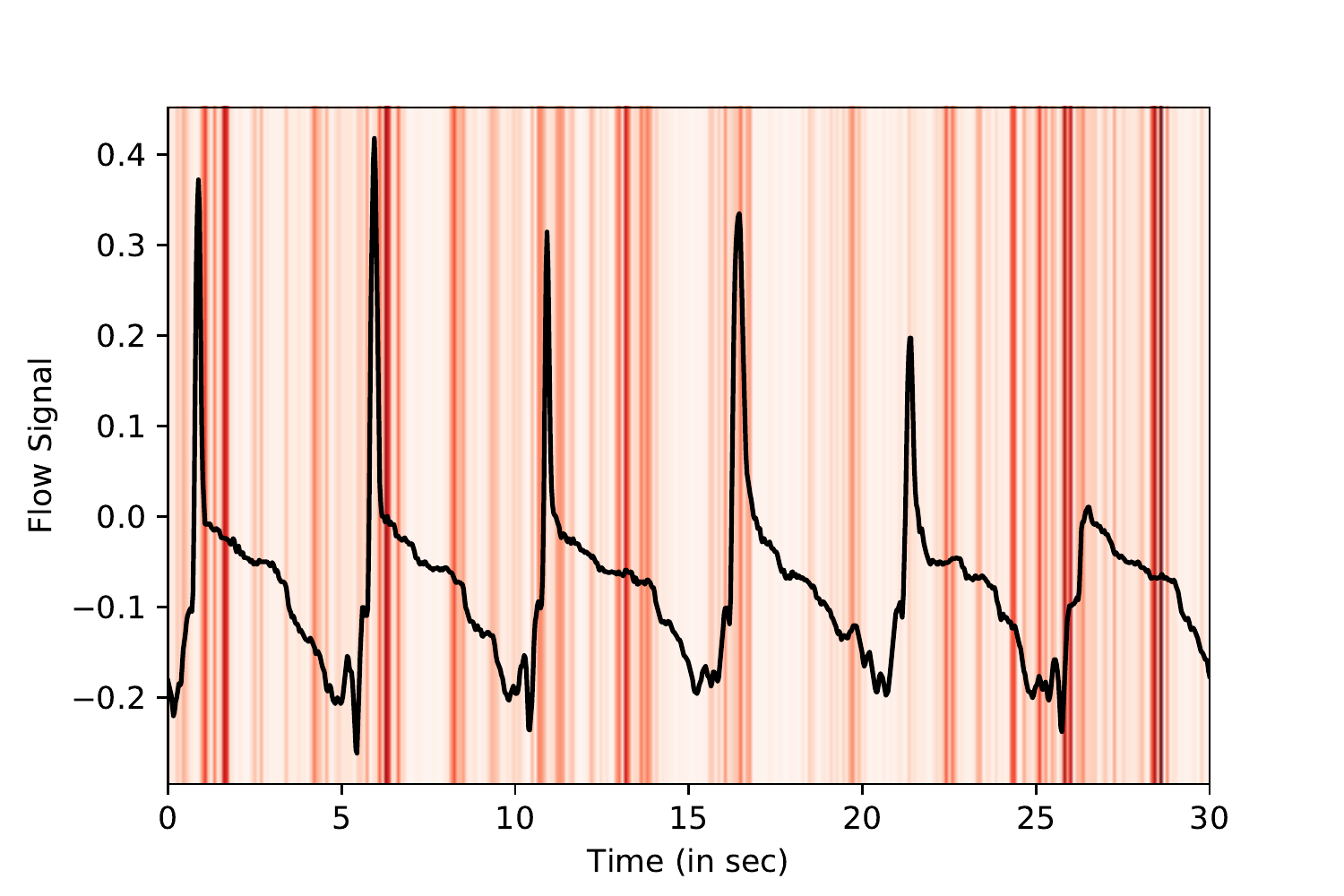}
\caption{Deep}
\label{fig:deep_sal}
\end{subfigure}

\caption{Saliency maps for the sample 30-second epochs correctly predicted by the model --- one each for wake, REM, light and deep sleep stages. Darker shade reflects higher weighting by the model.}
\label{fig:saliency}
\vspace{-1.5em}
\end{figure*}

\subsection{Effect of Cost-Sensitive Training}

Since our data has under-represented REM and Deep sleep annotations, we used a cost-sensitive prior. We demonstrate the effect of this prior by showing the confusion matrices for the Neural CRF and the cost-sensitive Neural CRF models in Figures~\ref{fig:conf}-\ref{fig:conf_cs}. By adding cost sensitive prior, the class accuracies increase across the board, significantly for the under-represented classes of REM and Deep sleep. Hence, using a cost-sensitive prior for lifting the weights of the under-represented classes during training is helpful for reducing the effect of imbalanced class distributions.  

\subsection{Using Flow Signal vs Other Signal Sources}
Our flow signal based models are able to detect the wake state accurately and light sleep with good accuracy, but have real difficulty detecting the REM and deep sleep. This is also visible in the \textit{t-SNE}~\cite{maaten2008visualizing} plot in Figure~\ref{fig:tsne}, where we plot in 2D the activations of the GRU-RNN layer of our cost sensitive neural CRF for a representative sample of annotations.  

Other signals such as no-contact and chest band based signals~\cite{zhao2017learning} have relatively lower performance on the wake and deep sleep states detection while are able to detect the REM 
and light sleep states accurately. We can however, say that using flow signal for the sleep staging does better than actigraphy, since we are able to detect the light sleep quite accurately compared to the actigraphy, in addition to the wake sleep actigraphy has shown to detect accurately. However, making a direct comparison with these studies might be misplaced since they have used a young or healthy 
population of subjects, while our population consists of sleep apnea patients. Previous attempts~\cite{redmond2006cardiorespiratory} on sleep apnea patients 
have observed lower accuracy compared to the healthy subjects since the sleep dynamics exhibited by sleep apnea patients are harder to predict than those of 
healthy subjects. Also, our study has a direct use case for the sleep apnea patients on CPAP treatment. Given the results on sleep efficiency task, flow signal can be used for monitoring the sleep efficiency of the patients, which is a very useful metric for the success of the therapy.

 %\begin{figure}[!b]
%\centering
%\includegraphics[width=0.3\textwidth]{figures/transition_m.pdf}
%\caption{Sleep Stage Transition matrix from the Regularized Cost-Sensitive Neural CRF model.}
%\label{fig:trans_crf}
%\end{figure}

\subsection{Flow Signal Saliency}

One of the most common criticisms of deep learning methods comes from the black-box nature 
of the models. We present the flow signal saliency as an exercise to interpret the model's basis for prediction.
In our case, the CRF layer's transition matrix (Figure \ref{fig:trans_crf}) helped us understand the output sleep 
stage dynamics learned by the model. In order to get an understanding of how the model is predicting the sleep stages from the input flow signal, we adopt the saliency map approach proposed by Simonayan et al.~\cite{simonyan2013deep} to interpret CNNs for image classification. The idea is to take the gradient of the classification scores 
with respect to the input image to learn weights of pixels the model is \emph{``looking" at} while making predictions. 

We use the same approach in our setting by learning the weights 
of model saliency over flow signal time-series by taking gradients of the classification scores with respect to 
the input flow signal. Figure~\ref{fig:saliency} shows one representative sample 30-second epoch (that was correctly classified by model) for each sleep stage, and their saliency weights over the time-series values. 

Through a visual inspection we can observe that the models seem to focus on two phases of the respiratory cycle, namely plateaus in flow closest to zero between inhalation and exhalation and periods of maximal change in flow rates.  Respiratory rate variability~\cite{gutierrez2016respiratory} and respiratory 
effort amplitude~\cite{douglas1982respiration} differ depending on stage of sleep.  Thus it is conceivable that the models may be extracting information that approximates respiratory physiology features in trying to classify sleep stage. On the other hand, the model's saliency map may also represent new and unknown phenomena that could be useful for medical researchers to investigate.

%% file: conclusions.tex
In this work, we present the first study on using flow signal for automated sleep staging. We utilize a neural CRF architecture that combines the representational power of deep neural networks with the modeling strength of structured output models to get the best of both worlds. For our neural model, we employ a deep CNN to learn high-level informative features from flow signals. A GRU-based RNN is used to encode features for classification by modeling temporal contextual information. The CRF jointly models the output sequence to capture temporal dynamics in the sleep states. Domain-dependent priors were used to regularize the network.

%Our contributions are two-fold: 1) We present the first study on performing sleep staging on flow data that has a readily available application in 
%sleep apnea therapy using CPAP, and 2) We present a neural condition random field model that jointly captures the temporal dynamics of sleep states and the input 
%signal features.  

%While the existing works in the area have entirely focused on feature extraction from the input signal; we present a method to exploit the temporal dynamics in the output sleep states. Existing models exploit the temporal dynamics only in the input space by using recurrent architectures - completely ignoring the temporal dynamics in the sleep states. In order to achieve this, we use the structured conditional approach by augmenting 
%the  deep neural network with conditional random field for end-to-end training. 

Our method substantially improves the classification performance over the baseline deep learning methods. We further demonstrate that using cost-sensitive prior for tackling class imbalance and sparse  
regularization on weights further improves the model performance. Our neural model (R-CNN) and the CRF approach can be augmented to the existing methods for improved sleep staging.
In terms of implications for the sleep care, our method has an existing and immediate use case as it can be employed to track the response of patients to 
the CPAP therapy by automatically and accurately tracking sleep stages and overall sleep quality. Hence, our study is helpful in advancing clinical sleep 
research and motivates researchers to investigate the effect of CPAP on sleep architecture of subjects.

%In the future, we would like to apply similar approaches to a multiple-signal data set.  While a single flow signal is available from standard PSG data as an analog to CPAP airflow, CPAP devices provide concomitant flow, pressure and leak signals with a more feature-rich data set.  We would like to apply similar methods to develop models to predict sleep stages based upon multiple signals.  %We could then expand to real-world comparisons of model outputs versus subjective experiences.}